\documentclass[english]{revtex4}
\usepackage[T1]{}
\usepackage[latin9]{inputenc}
\usepackage{amsmath}
\usepackage{graphicx}
\usepackage{amssymb}
\usepackage{esint}

\makeatletter
\usepackage{babel}
\makeatother

\begin{document}

\title{Holographic description of asymptotically AdS$_{2}$ collapse geometries}

\author{David A. Lowe and Shubho Roy}

\email{lowe@brown.edu, sroy@het.brown.edu}

\affiliation{Department of Physics, Brown University, Providence, RI 02912, USA}

\begin{abstract}
The mapping between bulk supergravity fields in anti-de Sitter space
and operators in the dual boundary conformal field theory usually
relies heavily on the available global symmetries. In the present
work, we study a generalization of this mapping to time dependent
situations, for the simple case of collapsing shock waves in two spacetime
dimensions. The construction makes use of analyticity of the conformal
field theory and the properties of the asymptotic bulk geometry to
reconstruct the non-analytic bulk observables. Many of the features
of this construction are expected to apply to higher dimensional asymptotically
anti-de Sitter spacetimes and their conformal field theory duals.
\end{abstract}
\maketitle

\section{Introduction}

In a series of papers \cite{Hamilton:2005ju,Hamilton:2006az,Hamilton:2006fh}
a reformulation of the Lorentzian AdS$_{D}$/CFT$_{D-1}$ correspondence
\cite{Maldacena:1997re,Gubser:1998bc,Witten:1998qj,Aharony:1999ti,Balasubramanian:1998sn}
was worked out in the leading semiclassical ($N\rightarrow\infty$)
approximation. This reformulation was based on mapping normalizable
bulk fields, (on the boundary $\phi(z,x)\sim z^{\Delta}\phi_{0}(x)$
as $z\to0$) to local CFT operators $\mathcal{O}(x)$ \cite{Klebanov:1999tb}
\[
\phi_{0}(x)\leftrightarrow\mathcal{O}(x)_{CFT}\,.\]
Here $z\rightarrow0$ on the boundary and $x$ coordinatizes the boundary.
The central aim of this reformulation was to recover approximate locality
in the bulk in the most transparent manner - by mapping on-shell bulk
insertions to a delocalized boundary (CFT) operator with compact support
on the boundary,

\[
\phi(z,x)\leftrightarrow\int dx'\, K(x'|x,z)\mathcal{O}(x')_{CFT}\,.\]

This was an improvement over earlier attempts \cite{Banks:1998dd,Balasubramanian:1999ri,Bena:1999jv}
which generally involved representation of a local bulk insertion
in terms of a nonlocal CFT operator with support over the $\emph{entire}$
boundary and hence required delicate cancellations to recover bulk
locality. This boundary-to-bulk map or the $\emph{smearing}$ $\emph{function}$,
$K(x'|x,z)$ constructed for various coordinate systems was nonvanishing
only for points on the boundary spacelike separated from the local
bulk insertion. The smearing function immediately reproduces the bulk
correlators in terms of the boundary correlators, for example

\[
\left\langle \phi(x_{1},z_{1})\phi(x_{2},z_{2})\right\rangle =\int dx_{1}'dx_{2}'K(x_{1}'|x_{1},z_{1})K(x_{2}'|x_{2},z_{2})\left\langle \mathcal{O}(x_{1}')\mathcal{O}(x_{2}')\right\rangle _{CFT}\,.\]

In the case of accelerating Rindler coordinates in AdS$_{3}$, the
smearing function could be expressed as a function with support on
a disc on the $\emph{complexified}$ boundary \cite{Hamilton:2006fh}.
This yielded a more transparent accounting of holographic entropy.
The BTZ black hole \cite{Banados:1992wn}, which can be conveniently
obtained as a periodic identification of the AdS$_{3}$ Rindler coordinates
was then considered. It was shown that for local bulk fields inside
the horizon one needed smeared CFT operators on the boundaries of
both the left and right Rindler wedges (see also \cite{Kraus:2002iv}).
However, even in these cases, using analytic continuation, one could
reduce to CFT operators smeared over a single complexified boundary
\cite{Hamilton:2006fh}. 

In the present paper we generalize these results to an asymptotically
AdS$_{2}$ spacetime with a null collapsing shockwave. The outline
of this paper is as follows. In section \ref{sec:the-ads_{2}-vaidya}
we review the coordinate system(s) employed for this Vaidya spacetime
and provide the Penrose diagram. We begin by constructing the boundary
to bulk map for points outside the black hole horizon in section \ref{sec:Points-Outside-the}.
First, in section \ref{sub:Zero-Mass} we consider a massless scalar,
where the result is identical to the pure AdS$_{2}$ case when written
using global null coordinates, despite the time-dependent geometry.
In section \ref{sub:Non-vanishing-mass} we deal with the slightly
more complicated massive case where we need to propagate the field
across the shock using appropriate matching conditions. The smearing
function again is only nonvanishing in the spacelike separated region
on the boundary. 

The results are extended to points inside the horizon in section \ref{sec:Points-inside-the}.
It is still possible to express local bulk fields in terms of CFT
operators on the single boundary at infinity, provided these are analytically
continued to complex values of the boundary coordinates. Thus we accomplish
the aim of completely characterizing a local theory propagating in
a $\emph{time-dependent}$ bulk geometry in terms of the analytic
boundary theory.

\section{the ads$_{2}$ vaidya spacetime\label{sec:the-ads_{2}-vaidya}}

We consider a 2d black hole which can be obtained by a dimensional
reduction of a BTZ formed by the gravitational collapse of a null
dust in 3d asymptotically AdS spacetime \cite{Maeda:2006cy,Husain:1994xa}
given by the metric,

\begin{equation}
ds^{2}=-f(r)dv^{2}+2dvdr=-f(r)dt^{2}+\frac{dr^{2}}{f(r)}\ \qquad(r>0)\,,\label{eq:vaidya}\end{equation}
where

\[
dt=dv-\frac{dr}{f(r)}\,,\]
and

\[
f(r)=\begin{array}{cc}
r^{2}-r_{0}^{2}, & v\geq0\\
1+r^{2}, & v<0\end{array}\,.\]
The two-dimensional action is the Jackiw-Teitelboim gravity theory
\cite{Jackiw1984,Teitelboim1984}. Such a black hole is formed by
doing cylindrical reduction on a spacetime with a negative cosmological
constant $\Lambda$ as a result of the gravitational collapse of a
null dust shell. We set $\Lambda=-1$ from now on. 

\begin{figure}
\includegraphics{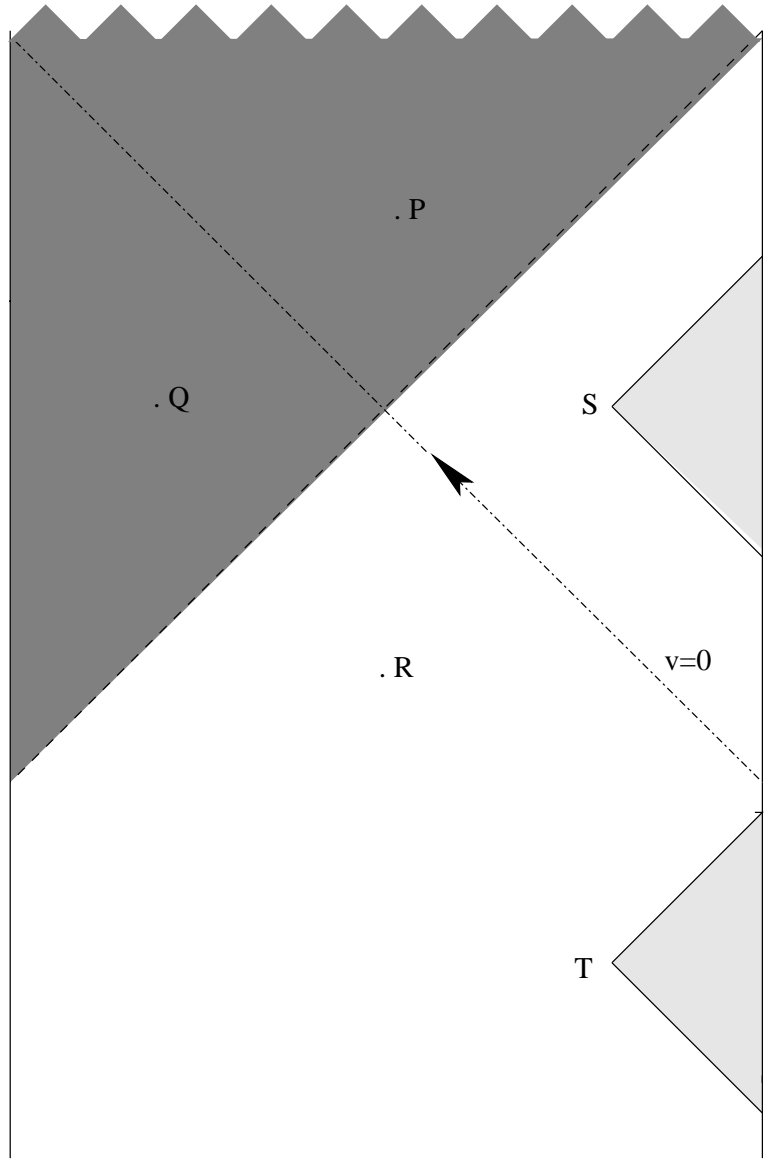}

\caption{\label{fig:Penrose-diagram}Penrose diagram}

\end{figure}

Tortoise coordinates $(t,r_{*})$ are defined as,

\begin{equation}
t=v-\int\frac{dr}{f(r)}\label{eq:tortoise}\end{equation}
and

\[
r_{*}=\int_{\infty}^{r}\frac{dr}{f(r)}=\begin{array}{cc}
\frac{1}{2r_{0}}ln\big|\frac{r-r_{0}}{r+r_{0}}\big|, & v\geq0\\
\tan^{-1}r-\pi/2 & v<0\end{array}\,.\]
Here the constant of integration $-\pi/2$ is chosen to make $r_{*}\rightarrow0$
as $r\rightarrow\infty$ from both sides of the shock. In order to
distinguish between the local coordinates defined in the regions $v\geq0$
and $(v<0)$, we shall use the superscript $+$ and $-$ respectively.
The range of the local coordinates are $-\infty<r_{*}^{+}\leq0$ for
points outside the horizon $(r>0)$ and $-\pi/2<r_{*}^{-}<0$. Inverting
the above relation, we have $r_{*}^{\pm}$ in terms of $r$,

\begin{equation}
r=-r_{0}\coth(r_{*}^{+}r_{0})=-\cot r_{*}^{-}.\label{eq:tortrel}\end{equation}
The Penrose diagram is shown in figure \ref{fig:Penrose-diagram}.%

\section{Boundary/Bulk duality outside the horizon\label{sec:Points-Outside-the}}

\subsection{Zero Mass \label{sub:Zero-Mass}}

A free massless scalar minimally coupled to Jackiw-Teitelboim gravity,
propagating in such a black hole spacetime satisfies the equation
of motion,

\[
\frac{1}{\sqrt{-g}}\partial_{a}(\sqrt{-g}g^{ab}\partial_{b}\phi)=0\,.\]

\subsubsection{Smearing function for point having support to either the past or
future of the shock }

In tortoise coordinates (\ref{eq:tortoise}) the wave equation is
simply 

\[
(-\partial_{t}^{2}+\partial_{r_{*}^{\pm}}^{2})\phi=0\,.\]
Now Green's theorem is,

\[
\phi(x)=\int dS'\sqrt{h}\eta^{a}\left(\partial'_{a}\phi(x')G(x,x')-\phi(x')\partial'_{a}G(x,x')\right)\]
where $h$ is the induced metric and $\eta$ is the normal derivative
on the boundary at infinity. We begin by considering points such as
S and T in figure \ref{fig:Penrose-diagram}. If we can find a Green's
function $G(r'_{*}-r_{*},t'-t)$ which has support inside the spacelike
cone extending from the bulk point to the boundary, e.g.

\[
G\varpropto\theta(r'_{*}-r_{*}-|t-t'|)\theta(r'_{*}-r_{*})\,,\]
 then we can extract the desired smearing function. We obtain such
a Green's function using contour integration,

\begin{equation}
(-\partial_{t}^{2}+\partial_{r_{*}}^{2})G(r'_{*}-r_{*},t'-t)=\delta(r_{*}'-r_{*})\delta(t'-t)\,.\label{eq:green}\end{equation}
In momentum space,

\[
(\omega^{2}-k^{2})G(k,\omega)=1\,.\]
So formally,

\[
G(r'_{*}-r_{*},t'-t)=\int\frac{d\omega dk}{(2\pi)^{2}}\frac{1}{\omega^{2}-k^{2}}e^{i(k\Delta r_{*}-\omega\Delta t)}\]
where $\Delta r_{*}=r'_{*}-r_{*}$, $\Delta t=t'-t$. Now lets do
the $k$ integration first by closing the contour in the upper half-plane,

\begin{eqnarray*}
G(r'_{*}-r_{*},t'-t) & = & -\int\frac{d\omega}{2\pi}e^{-i\omega\Delta t}\int\frac{dk}{2\pi}\frac{1}{(k-i\epsilon)^{2}-\omega^{2}}e^{ik\Delta r_{*}}\\
 & = & \frac{1}{i}\theta(\Delta r_{*})\int\frac{d\omega}{2\pi}e^{-i\omega\Delta t}\frac{e^{i\omega\Delta r_{*}}-e^{-i\omega\Delta r_{*}}}{2\omega}\\
 & = & \frac{1}{4}\theta(\Delta r_{*})\left(\mbox{sgn}(\Delta r_{*}+\Delta t)+\mbox{sgn}(\Delta r_{*}-\Delta t)\right)\\
 & = & \frac{1}{2}\theta(\Delta r_{*})\theta(\Delta r_{*}-|\Delta t|)\,.\end{eqnarray*}
Now the normalizable modes have the asymptotic fall off,

\[
\phi(r_{*},t)\sim r_{*}\phi_{0}(t)\]
as $r_{*}\to0$ since for a massless minimally coupled scalar the
scaling dimension is $\Delta=\frac{1}{2}+\sqrt{\frac{1}{4}+m^{2}R^{2}}=1$.
So according to to Green's theorem,

\[
\phi(r_{*},t)=\int dt'\phi_{0}(t')G(0-r_{*},t'-t)=\frac{1}{2}\int_{t+r_{*}}^{t-r_{*}}dt'\phi_{0}(t')\]
(recall that $r_{*}<0$) and hence the smearing function is simply
the Green's function with compact support on spacelike separated region
on the boundary.

\subsubsection{Smearing function for point with support to both past and future
of shock}

\begin{figure}
\includegraphics{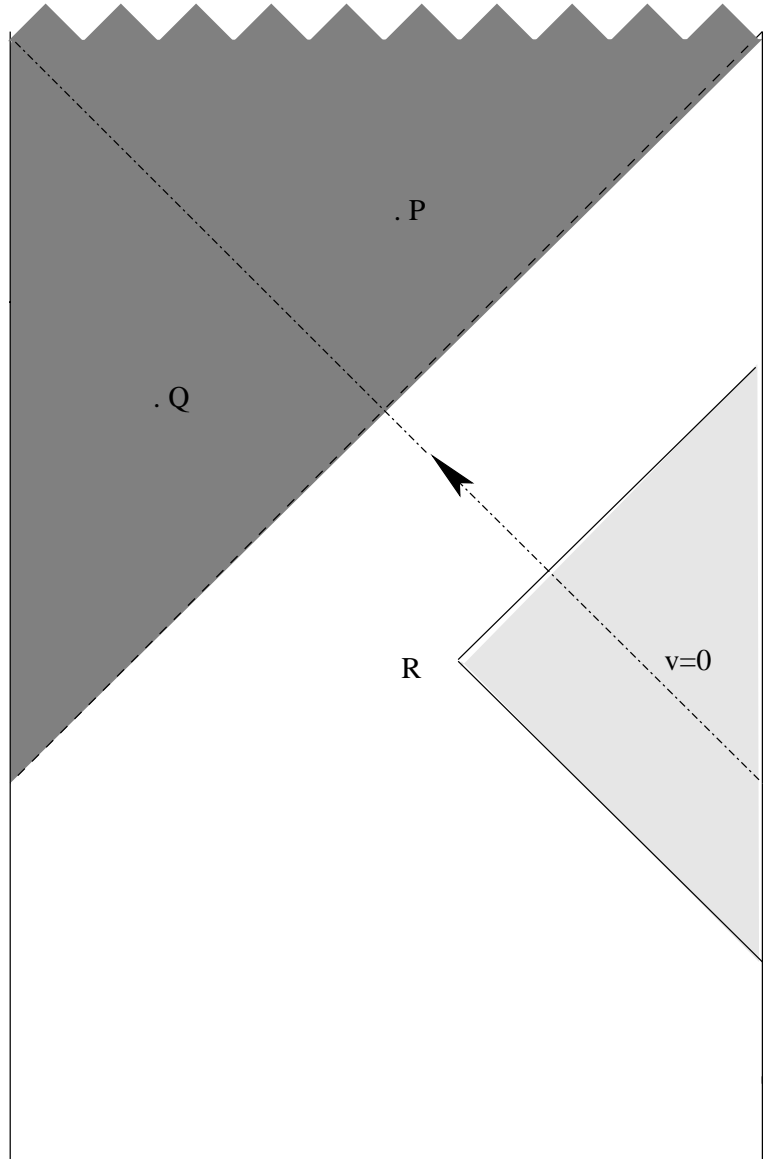}

\caption{Penrose diagram showing the a bulk field reconstructed using CFT operators
with support on the past and future of the shockwave.\label{fig:Penrose-diagram-showing}}

\end{figure}

In this section we consider points such as R from figure \ref{fig:Penrose-diagram-showing}.
For convience we will use null coordinates for the calculation. For
a null ray

\begin{eqnarray*}
2dr/f(r)-dv & = & 0\end{eqnarray*}
Introduce new coordinate $u$ 

\[
2r_{*}-v=u\]
where $u$ is a constant over the ray. At the shock $v=0$, we have
the relation

\[
u=2r_{*}\]
So according to (\ref{eq:tortrel}) the values of $u$ (on the same
null ray) across the shock are related by,

\[
r_{0}\coth(u^{+}r_{0}/2)=\cot(u^{-}/2)\]
 or,\[
u^{-}=h(u^{+})=2\cot^{-1}(r_{0}\coth(u^{+}r_{0}/2))\,.\]
So now we can make a ${\bf global}$ null coordinate,\[
u=\left\{ \begin{array}{c}
u^{-},\qquad v<0\\
h(u^{+}),\: v>0\end{array}\right.\,.\]
In $u$,$v$ parameters, the metric is,

\[
ds^{2}=F(u,v)dudv\]
where the conformal factor is,

\[
F(u,v)=\left\{ \begin{array}{cc}
f\left(r(u,v)\right)=\csc^{2}\left((u+v)/2\right), & v<0\\
f\left(r(h^{-1}(u),v)\right)\frac{dh^{-1}(u)}{du}=\frac{2r_{0}^{2}\csc h^{2}\left(\coth^{-1}\left(\left(\cot(u/2)\right)/r_{0}\right)+vr_{0}/2\right)}{(1+r_{0}^{2})\cos u-(1-r_{0}^{2})}, & v>0\end{array}\right.\,.\]
Spacelike infinity $r_{*}=0$ is where

\[
u^{\pm}+v=\theta(v)h(u)+\theta(-v)u+v=0\,.\]
In the global $u,v$ parameters, the Green's function equation for
a massless minimally coupled scalar is,

\[
2\partial_{u}\partial_{v}G(u-u',v-v')=\delta(u-u')\delta(v-v')\]
So, we obtain

\[
G(u-u',v-v')=\frac{1}{2}\theta(v'-v)\theta(u'-u)\,.\]
This is a ${\bf global}$ Green's function since the coordinate $u$
is continuous across the shock. Again this Green's function is nonvanishing
only in the spacelike separated region on the right boundary and hence
furnishes a smearing function which has a compact support on the timelike
boundary at infinity,

\begin{eqnarray*}
\phi(u,v) & = & \frac{1}{2}\int_{-\infty}^{0}dv'\left(G(u-u',v-v')\partial_{u'+v'}\phi(u',v')\right)|_{u'=-v'}+\frac{1}{2}\int_{0}^{\infty}dv'\left(\frac{dh}{du}(v')\right)^{1/2}\left(G(u-u',v-v')\partial_{h(u')+v'}\phi(u',v')\right)|_{h(u')=-v'}\\
 & = & \int_{-\infty}^{\infty}dv'\frac{1}{2}\left(\theta(-v')\theta(u+v')\theta(v-v')+\theta(v')\theta(u-h^{-1}(-v'))\theta(v-v')\right)\left(\frac{dh}{du}(v')\right)^{1/2}\phi_{0}(v')\,.\end{eqnarray*}

Recall that normalizable fields have asymptotic behavior $\phi(u,v)\sim(u+v)\phi_{0}(v)$
where $\lim_{u\rightarrow-v}\phi_{0}(v)\neq0$ for $v<0$ while for
$v>0$, we have $\phi\sim(h(u)+v)\phi_{0}(v)$ with $\lim_{h(u)\rightarrow-v}\phi_{0}(v)\neq0$
again. So the smearing function for massless fields in Kruskal coordinates
is,

\[
K\left(v'|u,v\right)=\frac{1}{2}\left(\theta(-v')\theta(u+v')\theta(v-v')+\theta(v')\left(\frac{dh}{du}(v')\right)^{1/2}\theta(u-h^{-1}(-v'))\theta(v-v')\right)\,.\]
The expression clearly demonstrates the fact that the smearing function
cannot be expressed in terms of a AdS$_{2}$-covariant distance function
since the metric is discontinuous (non-analytic) across the shock
and cannot be covered by a single analytic coordinate patch although
locally it is pure AdS$_{2}$ everywhere. The $u^{\pm},v$ coordinates
parametrize null geodesics and $u^{\pm}$ suffers a jump across the
shock. As a result, in terms of these coordinates the smearing function
has a discontinuity across the shock. Nevertheless the smearing function
has support only for points on the right boundary spacelike separated
from the bulk point, showing the construction of \cite{Hamilton:2005ju,Hamilton:2006az,Hamilton:2006fh}
generalizes to time-dependent bulk geometries.

\subsection{Non-vanishing mass \label{sub:Non-vanishing-mass}}

The massless minimally coupled field is insensitive to the conformal
factor in the metric in the $u,v$ coordinates which leads to a very
simple Green's function in these coordinates. On the other hand, the
massive scalar does feel the conformal factor,

\[
\frac{1}{\sqrt{-g}}\partial_{a}(\sqrt{-g}g^{ab}\partial_{b}G(x,x'))-m^{2}G(x,x')=\frac{1}{\sqrt{-g}}\delta(x-x')\,.\]
Now, lets go back to local parameters $r_{*}^{\pm}$, $t^{\pm}$ since
that is more suitable for calculation,

\[
ds^{2}=\left\{ \begin{array}{cc}
\csc^{2}r_{*}^{-}(-dt^{-2}+dr_{*}^{-2}), & v<0\\
r_{0}^{2}\csc h^{2}(r_{*}^{+}r_{0})(-dt^{+2}+dr_{*}^{+2}), & v>0\end{array}\right.\]
and are related by continuity across the surface of the shock by,

\begin{eqnarray}
r & = & -r_{0}\coth(r_{*}^{+}r_{0})=-\cot r_{*}^{-}\nonumber \\
t^{+} & = & v-r_{*}^{+}(r_{*}^{-})=t^{-}+r_{*}^{-}-r_{*}^{+}(r_{*}^{-})\,.\label{eq:shockcoords}\end{eqnarray}

The Green's function for the $v<0$ region with support over the spacelike
separated boundary region is the same as the one for pure AdS$_{2}$
with support over space like separated region on the right boundary
and was worked out in\cite{Hamilton:2005ju} for arbitrary non-vanishing
mass,

\[
G^{-}(r_{*}'^{-}-r_{*}^{-},t'^{-}-t^{-})=\frac{1}{2}P_{\Delta-1}(\sigma)\theta(r_{*}'^{-}-r_{*}^{-})\theta(r_{*}'^{-}-r_{*}^{-}-|t'^{-}-t^{-}|)\]
where $\Delta=\frac{1}{2}+\sqrt{m^{2}+\frac{1}{4}}$ and the AdS$_{2}$
covariant distance function,

\begin{equation}
\sigma(x,x')=\frac{\cos(t'^{-}-t^{-})-\cos r_{*}^{-}\cos r_{*}'^{-}}{\sin r_{*}^{-}\sin r_{*}'^{-}}\label{eq:invdist}\end{equation}
and the $P_{\Delta-1}(\sigma)$ is the Legendre function. 

For the $v>0$ region, the relevant Green's function is the one for
a point outside the horizon of a BTZ black hole with support on the
spacelike region of the right boundary. This is again just the pure
AdS$_{2}$ global Green's function transformed to the set of local
coordinates\cite{Hamilton:2005ju},

\[
G^{+}(x,x')=\frac{1}{2}P_{\Delta-1}(\sigma)\theta(\rho'-\rho)\theta(\rho'-\rho-|\tau-\tau'|)\]
where $\rho$,$\tau$ coordinates are defined in terms of the old
$r,t^{+}$ by,

\begin{eqnarray}
\frac{\cos\tau}{\sin\rho} & =- & \frac{r}{r_{0}}\nonumber \\
\frac{\sin\tau}{\cos\rho} & = & \tanh t^{+}r_{0}\label{eq:rhotau}\end{eqnarray}
and the metric is then,

\[
ds^{2}=\csc^{2}\rho(-d\tau^{2}+d\rho^{2})\,.\]
In these coordinates we define the boundary field $\phi_{0}(\tau)$,

\[
\phi_{0}(\tau)=\lim_{\rho\rightarrow0}\frac{\phi(\tau,\rho)}{\sin^{\Delta}\rho}\]
and for the $v<0$ region,

\[
\phi_{0}(t^{-})=\lim_{r_{*}^{-}\rightarrow0}\frac{\phi(t^{-},r_{*}^{-})}{\sin^{\Delta}r_{*}^{-}}\,.\]

\subsubsection{Matching conditions across the shock}

\begin{itemize}
\item Continuity of the field
\end{itemize}
The field is continuous across the shock,

\[
\phi(v=0^{+},r)=\phi(v=0^{-},r).\]

\begin{itemize}
\item Continuity of the normal derivative of the field
\end{itemize}
By integrating the d'Alembertian over a Gaussian surface straddling
$v=0$ and noting that there are no sources of the $\phi$-field on
$v=0$,

\begin{eqnarray*}
\int d^{2}x\sqrt{-g}\left(\frac{1}{\sqrt{-g}}\partial_{a}(\sqrt{-g}g^{ab}\partial_{b}\phi)-m^{2}\phi\right) & = & 0\\
\eta_{a}(\sqrt{-g}g^{ab}\partial_{b}\phi(x))|^{v=\epsilon}- & \eta_{a}(\sqrt{-g}g^{ab}\partial_{b}\phi(x))|^{v=-\epsilon}= & 0\end{eqnarray*}
where $\eta_{a}$ is the outward unit normal vector to $v=$ const..
In the global coordinates $(r,v)$, the normal derivative is just
$\partial_{r}\phi$ and then the condition is simply,

\[
\partial_{r}\phi(v=0^{+},r)=\partial_{r}\phi(v=0^{-},r)\,.\]

\subsubsection{Smearing function for a point with its right spacelike separated
regions extending to either sides of the shock}

In this case we have a local bulk insertion for a point like R in
figure \ref{fig:Penrose-diagram-showing}. Our strategy is to first
use Green's theorem to express the normalizable bulk field at $x^{-}$
(the superscript indicates it is in the $v<0$ region) in terms of
an integral over spacelike infinity of the $v<0$ region $\emph{and}$
the surface of the shock $v=0$, which bounds the spacelike separated
region to the right of $x^{-}$

\begin{eqnarray*}
\phi(x^{-}) & = & \int dt'^{-}\left((2\Delta-1)G^{-}(x^{-},x'^{-})\sin^{\Delta-1}r_{*}'^{-}\phi_{0}(t'^{-})\right)|_{r_{*}'^{-}\rightarrow0}\\
 &  & \qquad\qquad\qquad\qquad+\int dr_{*}'^{-}\left(\phi(x'^{-})\left(\overleftrightarrow{\partial_{r_{*}'^{-}}}-\overleftrightarrow{\partial_{t'^{-}}}\right)G^{-}(x^{-},x'^{-})\right)_{r_{*}'^{-}+t'^{-}=0}\,.\end{eqnarray*}
Then using the matching conditions across the shock we convert the
integral over the shock to an integral over the timelike infinity
of the $v>0$ region

\[
\int dr_{*}'^{-}\left(\phi(x'^{-})\left(\overleftrightarrow{\partial_{r_{*}'^{-}}}-\overleftrightarrow{\partial_{t'^{-}}}\right)G^{-}(x^{-},x'^{-})\right)_{r_{*}'^{-}+t'^{-}=0}\qquad\qquad\qquad\qquad\qquad\qquad\qquad\qquad\qquad\qquad\qquad\qquad\qquad\qquad\qquad\qquad\qquad\]

\[
\qquad\qquad\qquad=\int d\tau'\phi_{0}(\tau')\left(\int dr_{*}'^{-}K^{+}(\tau'|\rho,\tau)\left(\overleftrightarrow{\partial_{r_{*}'^{-}}}-\overleftrightarrow{\partial_{t'^{-}}}\right)G^{-}(x^{-},x'^{-})\right)\Bigl|_{\begin{array}{c}
\rho+\tau=r_{*}'^{-}+t'^{-}=0\\
\rho'=0\end{array}}\]
where $\rho=\rho(r_{*}'^{-},t'^{-}),$ $\tau=\tau(r_{*}'^{-},t'^{-})$
and $K^{+}$ is the smearing function for a field in the $v>0$ region.
So, the full smearing function is,

\begin{eqnarray}
K(x|x') & = & \theta(-t'^{-})(2\Delta-1)\left(G^{-}(x^{-},x'^{-})\sin^{\Delta-1}r_{*}'^{-}\right)|_{r_{*}'^{-}\rightarrow0}\nonumber \\
 &  & \,+\theta(\tau')\left(\int dr_{*}'^{-}K^{+}(\tau'|\rho,\tau)\left(\overleftrightarrow{\partial_{r_{*}'^{-}}}-\overleftrightarrow{\partial_{t'^{-}}}\right)G^{-}(x^{-},x'^{-})\right)\Bigl|_{\rho+\tau=r_{*}'^{-}+t'^{-}=0}\,.\label{eq:smearmass}\end{eqnarray}
The first term is simply the AdS$_{2}$ global smearing function as
worked out in \citep{Hamilton:2005ju},

\begin{eqnarray}
K_{1} & = & \theta(-t'^{-})(2\Delta-1)\left(G^{-}(x^{-},x'^{-})\sin^{\Delta-1}r_{*}'^{-}\right)|_{r_{*}'^{-}\rightarrow0}\nonumber \\
 & = & \theta(-t'^{-})\frac{2^{\Delta-1}\Gamma(\Delta+1/2)}{\sqrt{\pi}\Gamma(\Delta)}\left(\frac{\cos(t'^{-}-t^{-})-\cos r_{*}^{-}}{\sin r_{*}^{-}}\right)^{\Delta-1}\theta(-r_{*}^{-}-|t'^{-}-t^{-}|)\,.\label{eq:kone}\end{eqnarray}
The second term in (\ref{eq:smearmass}), the integral along the shock,
is a bit more involved. Explicitly (dropping the overall $\theta(\tau')$),

\begin{equation}
K_{2}=\int dr_{*}'^{-}K^{+}(\tau'|\rho,\tau)(\overleftrightarrow{\partial_{r_{*}'^{-}}}-\overleftrightarrow{\partial_{t}})G^{-}(r_{*}^{-},t^{-}|r_{*}'^{-},t'^{-})\label{eq:ktwo}\end{equation}
where

\[
K^{+}(\tau'|\rho,\tau)=\frac{2^{\Delta-1}\Gamma(\Delta+1/2)}{\sqrt{\pi}\Gamma(\Delta)}\left(\frac{\cos(\tau'-\tau)-\cos\rho}{\sin\rho}\right)^{\Delta-1}\theta(-\rho-|\tau'-\tau|)\]
and

\[
G^{-}(r_{*}'^{-}-r_{*}^{-},t'^{-}-t^{-})=\frac{1}{2}P_{\Delta-1}(\sigma)\theta(r_{*}'^{-}-r_{*}^{-})\theta(r_{*}'^{-}-r_{*}^{-}-|t'^{-}-t^{-}|)\]
where $\Delta=\frac{1}{2}+\sqrt{m^{2}+\frac{1}{4}}$ and the AdS$_{2}$
covariant distance function is given in (\ref{eq:invdist}). $K_{2}$
is more conveniently expressed as a sum of two contributions

\[
K_{2}=K_{2I}+K_{2II}\]
where

\begin{eqnarray*}
K_{2I} & = & \theta(\tau')\int dr_{*}'^{-}K^{+}(\tau'|\rho,\tau=-\rho)(\partial_{r_{*}'^{-}}-\partial_{t})G^{-}(r_{*}^{-},t^{-}|r_{*}'^{-},t'^{-})|_{t'^{-}=-r_{*}'^{-}}\\
 & = & \left(-2\sin\frac{\tau'}{2}\right)^{\Delta}\left(\frac{2^{\Delta-2}\Gamma(\Delta+1/2)}{\sqrt{\pi}\Gamma(\Delta-1)}\right)\\
 & \times & \int dr_{*}'^{-}\left(\left(\frac{\sin(\tau'/2+\rho)}{\sin\rho}\right)^{\Delta-1}\theta(-\rho-|\tau'+\rho|)\left(\frac{\cos r_{*}'^{-}-\cos t^{-}}{\sin r_{*}^{-}\sin^{2}r_{*}'^{-}}\right)\frac{\sigma P_{\Delta-1}-P_{\Delta-2}}{\sigma^{2}-1}\theta(\Delta r_{*}^{-})\theta(\Delta r_{*}^{-}-|r_{*}'^{-}-t^{-}|)\right)\\
 &  & \qquad\qquad\qquad\qquad\qquad\qquad\qquad+\frac{1}{2}\theta(\tau')\theta\left(-\frac{r_{*}^{-}+t^{-}}{2}\right)\left(K^{+}\left(\tau'|\rho,\tau=-\rho\right)P_{\Delta-1}\left(\sigma\right)\right)|_{r_{*}'^{-}=\frac{r_{*}^{-}-t^{-}}{2}=-t'^{-}}\end{eqnarray*}
while

\begin{eqnarray*}
K_{2II} & = & \int dr_{*}'^{-}\left(G^{-}(r_{*}^{-},t^{-}|r_{*}'^{-},t'^{-})(\partial_{r_{*}'^{-}}-\partial_{t^{-}})K_{1}(\rho,\tau|\tau')\right)|_{\begin{array}{c}
\tau=-\rho\\
t'^{-}=-r_{*}'^{-}\end{array}}\\
 & = & \left(-2\sin\frac{\tau'}{2}\right)^{\Delta}\left(\frac{2^{\Delta-3}\Gamma(\Delta+1/2)}{\sqrt{\pi}\Gamma(\Delta-1)}\right)\\
 & \times & \ \quad\int dr_{*}'^{-}P_{\Delta-1}(\sigma)\theta(\Delta r_{*}^{-})\theta(\Delta r_{*}^{-}-|-r_{*}'^{-}-t^{-}|)\frac{\sec^{2}r_{*}'^{-}}{1/r_{0}+r_{0}\tan^{2}r_{*}'^{-}}\frac{\sin^{\Delta-2}(\tau'+\rho/2)}{\sin^{\Delta}\rho}\theta(-\rho-|\tau'+\rho|)\end{eqnarray*}
where on the shock (\ref{eq:shockcoords}) reduces to

\[
\tan\rho=\tan\rho(r_{*}'^{-})=r_{0}\tan r_{*}'^{-}\,.\]
So, the full smearing function in this case looks complicated as the
bulk geometry changes across the shell from pure AdS$_{2}$ to a AdS$_{2}$
black hole, but nevertheless has support only on the right boundary
region spacelike separated from the bulk point.

\section{Points inside the horizon\label{sec:Points-inside-the}}

As has been pointed out in\cite{Hamilton:2005ju} for operator insertions
at points inside the future Rindler horizon in pure AdS$_{2}$, the
smearing function has support on both boundaries i.e. boundaries of
the left and right Rindler wedges. For integer conformal dimension
$\Delta$,

\begin{equation}
\phi(P)=\int_{-\infty}^{\infty}dt\, K_{Rindler}^{R}(t|P)\phi_{0}^{Rindler,R}(t)+(-)^{\Delta}K_{Rindler}^{L}(t|P)\phi_{0}^{Rindler,L}(t)\label{eq:insidehoriz}\end{equation}
where 

\begin{eqnarray*}
K_{Rindler}^{R}(t|P) & = & \frac{2^{\Delta-1}\Gamma(\Delta+1/2)}{\sqrt{\pi}\Gamma(\Delta)}\lim_{r\rightarrow\infty}\left(\frac{\sigma}{r}\right)^{\Delta-1}\theta(\sigma-1)\\
K_{Rindler}^{L}(t|P) & = & \frac{2^{\Delta-1}\Gamma(\Delta+1/2)}{\sqrt{\pi}\Gamma(\Delta)}\lim_{r\rightarrow\infty}\left(-\frac{\sigma}{r}\right)^{\Delta-1}\theta(-\sigma-1)\end{eqnarray*}
and $\sigma=\sigma(t,r|P)$ is the AdS$_{2}$ invariant distance.
\footnote{For non-integer $\Delta$, the result is a bit more complicated and
can be found in\cite{Hamilton:2005ju}. The upshot is that a local
bulk operator inside the black hole horizon is delocalized over both
the left and right boundaries.%
}

However the integral over the left Rindler boundary can be mapped
back to the right Rindler boundary using the identification,

\[
\phi_{0}^{Rindler,L}(t)=\phi_{0}^{Rindler,R}(t+i\pi)\,.\]
This identification was arrived at after noting that a complex change
of Rindler coordinates takes one from the left to the right Rindler
wedge,

\[
t\rightarrow t+i\pi.\]
So combining these information we have a smearing function which has
compact support entirely on the $\emph{complexified}$ right boundary. 

\begin{equation}
\phi(P)=\phi(P)=\int_{-\infty}^{\infty}dt\, K_{Rindler}^{R}(t|P)\phi_{0}^{Rindler,R}(t)+(-)^{\Delta}K_{Rindler}^{L}(t|P)\phi_{0}^{Rindler,R}(t+i\pi)\,.\label{eq:insideone}\end{equation}
A similar result for the BTZ black hole was also derived in \citep{Hamilton:2006fh}. 

Now let us extend these results to the shockwave geometry. In this
case we no longer have the same global isometries used in the derivation
of the expressions (\ref{eq:insidehoriz}) and (\ref{eq:insideone}).
However for bulk points to the future of the shell such as point P
in figure \ref{fig:Penrose-diagram}, by analytic continuation, we
can pretend to be in an eternal AdS$_{2}$ black hole and the expression
(\ref{eq:insideone}) carries over \cite{Hamilton:2006fh}.

New features appear when we study points to the past of the shock,
but inside the horizon, such as Q in figure \ref{fig:Penrose-diagram}.
For this point we first use the spacelike Green's function to express
the bulk scalar in terms of an integral on $v=0^{-}$ and the portion
of the boundary $r_{*}^{-}=0$ on the $v<0$ side

\[
\phi(Q)=\int dt'^{-}K_{1}(Q|t')\phi_{0}(t'^{-})+\int dr_{*}'^{-}\left(\phi(x'^{-})\left(\overleftrightarrow{\partial_{r_{*}'^{-}}}-\overleftrightarrow{\partial_{t'^{-}}}\right)G^{-}(Q,x'^{-})\right)_{r_{*}'^{-}+t'^{-}=0}\,.\]
We divide the integral on the shock into two parts, one outside and
the other inside the horizon $r=r_{0}$ ($r_{*}^{-}=\tanh^{-1}r_{0}-\pi/2$)

\begin{eqnarray*}
\phi(Q) & = & \int dt'^{-}K_{1}(Q|t')\phi_{0}(t'^{-})+\int dr_{*}'^{-}\left(\phi(x'^{-})\left(\overleftrightarrow{\partial_{r_{*}'^{-}}}-\overleftrightarrow{\partial_{t'^{-}}}\right)G^{-}(Q,x'^{-})\right)_{v=0,r<r_{0}}\\
 &  & \qquad\qquad\qquad\qquad\qquad\qquad\qquad+\int dr_{*}'^{-}\left(\phi(x'^{-})\left(\overleftrightarrow{\partial_{r_{*}'^{-}}}-\overleftrightarrow{\partial_{t'^{-}}}\right)G^{-}(Q,x'^{-})\right)_{v=0,r>r_{0}}\end{eqnarray*}
where $K_{1}$ is same as in (\ref{eq:kone}). Now we use the matching
conditions to express the integral over the shock $\emph{outside}$
$\emph{the}$ $horizon$ to an integral over the boundary $r_{*}^{+}=0$
just as was done for the point R. Finally for the integral over the
$v=0^{-}$ surface inside the horizon ($r<r_{0}$) side we again first
match it to quantities just across the shell i.e. $v=0^{+}$ and then
express it in terms of an integral over analytically continued (complexified)
right boundary just as done for the point P. The final expression
is,

\begin{eqnarray}
\phi(Q) & = & \int dt'^{-}K_{1}(Q|t')\phi_{0}(t'^{-})+\int d\tau'K_{2}(Q|\tau')\phi_{0}(\tau')\nonumber \\
 &  & \qquad+\int d\tau'K_{3I}(Q,\tau')\phi_{0}^{Rindler,R}(\tau')+\int d\tau'K_{3II}(Q,\tau')\phi_{0}^{Rindler,R}(\tau'+i\pi)\label{eq:finalans}\end{eqnarray}
where $K_{2}$ is given by (\ref{eq:ktwo}) while $K_{3}$'s are 

\begin{eqnarray*}
K_{3I}(Q,\tau') & = & \int dr_{*}^{-}\,\left[K_{Rindler}^{R}(x'^{-}|\tau')\left(\overleftrightarrow{\partial_{r_{*}'^{-}}}-\overleftrightarrow{\partial_{t'^{-}}}\right)G^{-}(Q,x'^{-})\right]_{v=0,r>r_{0}}\\
K_{3II}(Q,\tau') & = & \int dr_{*}^{-}\left[(-)^{\Delta}K_{Rindler}^{L}(x'^{-}|\tau')\left(\overleftrightarrow{\partial_{r_{*}'^{-}}}-\overleftrightarrow{\partial_{t'^{-}}}\right)G^{-}(Q,x'^{-})\right]_{v=0,r>r_{0}}\,.\end{eqnarray*}

This is a highly nontrivial result. The bulk theory describes propagation
in a non-analytic geometry. The boundary CFT nevertheless inherits
analytic amplitudes from boundary locality. By exploiting the analyticity
of the boundary theory, the expression (\ref{eq:finalans}) shows
how propagation in the bulk geometry is reconstructed. Bulk non-analyticity
shows up only in the smearing function factors.

\section{Conclusions}

In this paper we looked at a model time-dependent black hole geometry
which has a $single$ asymptotic AdS$_{2}$ boundary and constructed
a bulk-boundary map which represents local on-shell bulk field operators
as non-local boundary operators with compact support but on the boundary.
We used analyticity of the boundary CFT as a tool to construct such
maps for points inside the black hole horizon. The smearing function
is very similar to the one for a pure AdS$_{2}$, for points outside
the horizon, the function has support over only spacelike separated
boundary points. For points inside the horizon the integral is delocalized
over the single boundary at infinity, and requires a complex time
contour.

One can obtain a more general time-varying geometry by collapsing
a series of such null shells. The present construction should generalize
straightforwardly to this case, using the general time dependent Vaidya
metric \cite{Maeda:2006cy}. It will also be interesting to generalize
this construction to AdS bubble solutions relevant to cosmology, such
as those studied in \cite{Alberghi:1999kd,Freivogel:2005qh,Lowe:2007ek}.

\begin{acknowledgments}
We thank Alex Hamilton, Daniel Kabat and Gilad Lifschytz for discussions.
This research is supported in part by DOE grant DE-FG02-91ER40688-Task
A.

\appendix
\section{Coordinate Systems}
\end{acknowledgments}
In this appendix we review the transformations between the different
coordinate systems used above. The coordinates $(r_{*}^{+},t^{+})$
cover a patch to the future of the shock as described in section \ref{sec:the-ads_{2}-vaidya}.
These are related to the $(r,v)$ coordinates of the Vaidya metric
(\ref{eq:vaidya}) by

\[
r=-r_{0}\coth(r_{*}^{+}r_{0})\,,\qquad v=t^{+}+r^{+}\,.\]
The coordinates $(r_{*}^{-},t^{-})$ cover a patch to the past of
the shock where\[
r=-\cot r_{*}^{-}\,,\qquad v=t^{-}+r^{-}\,.\]
Therefore on the shock $v=0$, we have 

\[
r_{*}^{+}|_{v=0}=\frac{1}{r_{0}}\coth^{-1}\left(\frac{\cot r_{*}^{-}|_{v=0}}{r_{0}}\right)\]

\[
t^{+}|_{v=0}=-\frac{1}{r_{0}}\coth^{-1}\left(\frac{\cot r_{*}^{-}|_{v=0}}{r_{0}}\right)\,.\]

In section \ref{sub:Non-vanishing-mass} we also introduce the coordinates
$(\rho,\tau)$ in (\ref{eq:rhotau}). These are related to $(r_{*}^{+},t^{+})$
by

\[
\rho=-\sin^{-1}\left(\frac{1}{\coth^{2}r_{0}r_{*}^{+}\cosh^{2}r_{0}t^{+}-\sinh^{2}r_{0}t^{+}}\right)^{1/2}\,,\qquad\tau=\sin^{-1}\left(\frac{1}{\cosh^{2}r_{0}r_{*}^{+}\coth^{2}r_{0}t^{+}-\sinh^{2}r_{0}r_{*}^{+}}\right)^{1/2}\,.\]
On the surface of the shock we have

\[
\rho|_{v=0}=-\sin^{-1}\left(\frac{1}{1+\coth^{2}r_{0}r_{*}^{+}}\right)^{1/2}=-\tau|_{v=0}\,.\]
Finally, for completeness, the relation between the $(r_{*}^{-},t^{-})$
and $(\rho,\tau)$ coordinates is

\[
\frac{\cos\tau}{\sin\rho}=-\frac{r}{r_{0}}=\frac{1}{r_{0}}\cot r_{*}^{-}\]
\[
\sin(\tau+\rho)=\tanh(r_{*}^{-}+t^{-})r_{0}\,.\]

\bibliographystyle{brownphys}
\bibliography{collapse1}

\end{document}